**Atomistic insight into the effects of solute and pressure on phase transformation in titanium alloys**

Huicong Chen [a,*], Chenwei Shao [a], Zhuocheng Xie [a], Jun Song [b], Yu Zou [a,*]

[a] Department of Materials Science and Engineering, University of Toronto, 184 College Street, Toronto, M5S 3E4, Canada
[b] Department of Mining and Materials Engineering, McGill University, 3610 Rue University, Montréal, H3A 0C5, Canada

*Corresponding authors:* Huicong Chen, Email address: uc.chen@utoronto.ca; Yu Zou, Email: mse.zou@utoronto.ca

# Abstract

The phase stability and transformation between hexagonal close-packed (hcp) α-phase and body-centered cubic (bcc) β-phase in titanium (Ti) alloys are critical to their mechanical properties and manufacturing processes for engineering applications. However, many factors, both intrinsic and extrinsic (e.g., solute elements and external pressures, respectively), may govern their phase transformations dynamically, which is crucial to the design of new Ti alloys with desirable properties. In this work, we study the effects of various solute elements and external hydrostatic pressures on the solid-state phase transformations in Ti alloys using density functional theory (DFT) and nudged elastic band (NEB) calculations. The results show that both alloying and applied pressure reduce transformation barriers, with Al and Mo being most effective under ambient conditions, while Nb, V, Zr, and Sn show enhanced transformation kinetics under stress. Solute-induced modifications to the local electronic structure and bonding environment, particularly under pressure, contribute to variations in phase stability. We identify a synergistic interaction between solute effects and external stress, which facilitates phase transitions that are unachievable under static conditions. These findings provide atomistic insights into the coupled chemical-mechanical mechanisms underlying phase transformations in Ti alloys with improved phase stability and mechanical performance.

# 1. Introduction

Titanium (Ti) alloys are extensively employed in the aerospace, energy, chemical, military, marine and biomedical industries, owing to their superior strength-to-weight ratios and corrosion resistance, outperforming materials such as steel in these aspects [1-4]. Commercial Ti alloys predominantly exhibit two phases: hexagonal close-packed (hcp) α and body-centered cubic (bcc) β. The α-phase alloys are known for their high strength, especially at high temperatures, while β-phase alloys are characterized by superior ductility and formability [5-10]. Consequently, controlling the phase transformation between α and β phases and tailoring their morphologies is a central strategy in the development of advanced Ti alloys. Dynamic transformation, defined as the phase change from a parent to a product phase under mechanical loadings at varying temperatures, occurs in a thermodynamically unstable state [11]. This transformation critically influences microstructural features, such as morphology, size, volume fraction, and compositional distribution, and the resulting mechanical properties of Ti alloys [12-14]. Moreover, the dynamic transformation is also essential to the processing conditions, such as temperature and cooling rates, pressure and strain rate, and alloying elements [9, 10, 15-19]. So far, most studies have focused on the thermodynamics and kinetics of equilibrium phase transformations in Ti alloys, as well as the resulting static microstructures under controlled heat treatment conditions [20-23]. However, there have been few studies focusing on the atomistic mechanisms and dynamic evolution of phase transformations under non-equilibrium conditions, such as during rapid deformation or thermo-mechanical processing. In particular, the interplay between mechanical stress and solute in driving the α-β transformation remains poorly understood at the atomic scale, as well as their combined role in facilitating or hindering phase transformation pathways during dynamic loading.

Alloying, as an intrinsic factor, in is a popular strategy to tune the phase transformations in Ti alloys because the stability of α or β phase is strongly influenced by the type and concentration of solute elements. For example, elements such as Al, C and N, known as α stabilizers, enhance the stability of α phase by raising the temperature above which the alloy is single-phase β [15, 24]. In contrast, elements such as V, Mo and Nb can stabilize the β phase even at room temperature by destabilizing the α phase. These so-called β-stabilizers can lower the β transus temperature, leading to the phase diagram to be either the isomorphous or the eutectoid-type [11, 25-27]. By altering the β transus temperature and the thermodynamic driving force for transformation, alloying elements can influence the kinetics of dynamic phase transformations, thus impacting grain size, phase fractions, and the mechanical properties of the alloy [8, 15, 19, 28]. Moreover, solute atoms can form fine precipitates during phase transformation, enhancing alloy strength by impeding dislocation motion [11, 29]. While the thermodynamic effects of alloying elements on equilibrium phase stability have been extensively studied, their quantitative impact on transformation pathways during dynamic loading remains largely unexplored. Pressure, as an extrinsic factor, also exerts a considerable impact on dynamic phase transformations [30-32]. Increased pressure raises the thermodynamic driving force for the β-α transformation, thus affecting the kinetics of phase nucleation and growth [33, 34]. In the absence of external strain or stress, the primary driving force for a phase transition is the Gibbs free energy difference between the two phases, which varies with temperature. However, during thermomechanical deformation, the applied stress can contribute an additional driving force for phase transformation. Moreover, the diffusion kinetics and spatial distribution of solute elements during deformation differ from those in an undeformed system, altering the equilibrium conditions between phases [33, 34]. Recent experimental studies showed that pressure could shift the α-β equilibrium, stabilizing the β phase at lower temperatures and under reduced strain rates [35, 36]. This pressure-induced stabilization enhances superplasticity and improves flow properties [36]. Moreover, recent observations of dynamic transformation from α to β under tensile deformation after long-time equilibration suggest that external stress may promote this transformation [14, 35]. Consequently, a comprehensive understanding of the mechanisms governing phase stability under external pressure, along with the multiscale influence of solute–matrix interactions on transformability, is essential for advancing the design and development of high-performance Ti alloys.

The orientation relationship (OR) between α and β phases generally follows the Burgers orientation: $(0001)_\alpha \parallel (110)_\beta$ and $[11\bar{2}0]_\alpha \parallel [111]_\beta$ [37]. Recently, Cayron [38, 39] proposed a unified crystallographic model to understand the phase transitions across the fcc–bcc–hcp systems. This theoretical model can directly predict the displacive pathway between different phases Considering the defects evolved in the phase interface, Benrahah *et al.* [40] qualitatively evaluate the barrier energy in the bainite phase transformation and the solute atoms like C and Mn would affect the kinetics of the transition. However, these mention models are incapable of evaluating the transformability and quantitatively elucidate the contribution from solute or external loading. Although many computational efforts, such as phase field model [41, 42], molecular dynamic (MD) simulations [43, 44] and first principle calculations [45, 46] have been used to investigate the phase transformation mechanisms and predict the stability of different phases, a comprehensive understanding of the atomistic processes involved remains elusive. Critical questions persist regarding how pressure alters local structural instabilities and transformation energy barriers under mechanical loading. Moreover, a quantitative understanding of the interplay between pressure and solute chemistry in governing phase stability and transformation kinetics under dynamic conditions is still lacking.

To address this knowledge gap, the present study employs atomistic modeling to systematically investigate the coupled effects of pressure and solute elements on the dynamic α-β phase transformation in Ti alloys. Therefore, in this study, we elucidate the combined effects of alloying elements and external pressures on the ability of α-β transition in Ti-based alloys using first-principles calculations based on DFT. Accordingly, the transformation pathway from α to β was constructed to achieve the entropy landscapes through the nudged elastic band (NEB) calculations. Then, the energy barriers were computed to quantitatively evaluate the contribution of different solutes and external pressure to the improvement of α-β transition. The density of states (DOS) calculations and Bader charge analysis are employed to explore the physical effects of different alloying elements and external pressure on the transition.

## 2. Methodology

To investigate the effect of solutes and pressures on the ability of the phase transformation from α to β, the potential transformation paths at constant pressures are calculated via a generalized solid-state nudged elastic band (G-SSNEB) method [47] using DFT. The G-SSNEB technique involves the unit-cell and atomic degrees of freedom and is often applied to determine reaction pathways of solid-solid phase transitions [47, 48]. The α to β transformation follows the Burgers orientation relationship as $\alpha(0001)||\beta(110)$ and $\alpha(11\bar{2}0)||\beta(111)$ [49]. According to the analytical work by Cayon [38, 39], the intermediate configurations can be introduced into the transformation pathway during the α to β transition (**Fig. 1**(a)). In current calculations, supercells of the hcp and bcc structures with 128 atoms were constructed. One Ti atom was substituted by a solute atom (**Fig. 1**(a)) to generate different alloy systems, and solutes Al, Zr, Sn, V, Mo and Nb are considered in this work.

All DFT calculations were performed using the Vienna ab-initio simulation package (VASP) [50] within the projector-augmented wave (PAW) method [51]. The exchange-correlation interaction was described by the generalized gradient approximation (GGA) using Perdew–Burke–Ernzerhof (PBE) based on plane-wave basis sets [52]. The selection of valence electron configuration follows the recommendation by VASP for relatively accurate calculations, including Ti: $3s^23p^63d^34s^1$, Al: $3s^23p^1$, Zr: $4s^24p^64d^35s^1$, Sn: $4p^{10}5d^25s^2$, V: $3s^23p^63d^44s^1$, Mo: $4s^24p^64d^35s^1$ and Nb: $4s^24p^64d^35s^1$. The plane wave basis

kinetic cut off was set to 450 eV, and a Γ-centered k-point mesh was used. The NEB calculations stopped when the total force of an atomic configuration and the total energy of the cell was less than 0.01 eV/Å and $10^{-5}$ eV/cell, respectively. Γ-centered k-point mesh of $5 \times 5 \times 6$ was chosen for all calculations. Eight images (**Fig. 1**(a)) were inserted between the initial hcp structure and final bcc structure for the NEB calculations at various hydrostatic pressures 0 GPa, 5 GPa, 10 GPa and 15 GPa. The activation enthalpy (energy barrier) is defined as the enthalpy difference between the transition state (the highest energy point in the reaction pathway) and the initial hcp phase. To understand the effect of solutes or pressure on phase stability, the density of states (DOS) was calculated to reveal the evolution of the local band structure. Additionally, calculations of electron charge density difference (ECDD) and Bader charge analysis (BCA) were performed to investigate local electron transfer. A finer k-point mesh of $7 \times 7 \times 7$ was used for performing the calculations of DOS, ECDD and BCA.

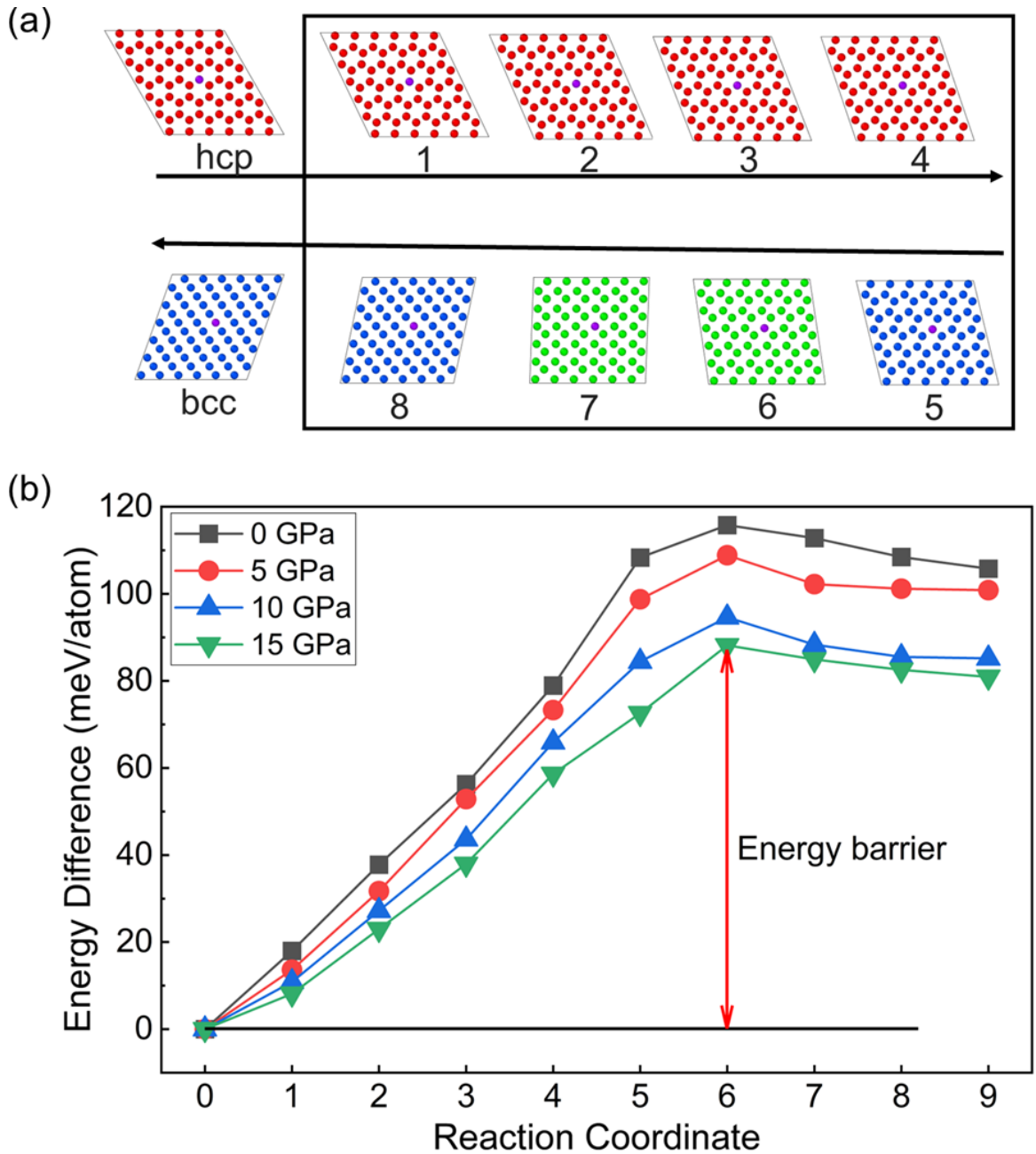

**Fig. 1.** Reaction pathway of phase transformation calculated by the G-SSNEB method with DFT. (a) 8 images were inserted between the initial hcp phase and the final bcc phase. Atom colors denote local crystal structures classified by common neighbour analysis (CNA) [53]: red (hcp), green (fcc), and blue (bcc). The purple atom represents a substitutional impurity. (b) Enthalpy landscape of the α-β phase transformation for the pure Ti system at 0 GPa, 5 GPa, 10 GPa, and 15 GPa.

# 3 Results

## 3.1 Transition pathway and energy barrier

**Fig. 2** shows the calculated phase enthalpy landscape from α to β in Ti-based alloys. The curves exhibit a characteristic shape where the entropy initially increases, reaches a highest point, and then decreases. This peak represents the energy barrier for the phase transformation between the hcp and bcc phases. With increasing pressure, the curves are lowered, and the peak height decreases, indicating a reduction in the energy barrier for the transformation. This suggests that higher pressure facilitates the phase transformation by making it thermodynamically and kinetically more favorable. Furthermore, the addition of a solute atom further modifies the entropy landscape. For example, when an Al (**Fig. 2**(a)) or V (**Fig. 2**(d)) atom is added, the peak height is slightly reduced compared to the trends for pure Ti (**Fig. 1**(b)), reflecting a lower energy barrier for the transformation.

To quantitatively evaluate how solute and pressure affect the transformability from hcp to bcc structures, the activation enthalpy was calculated, and the results are shown in **Fig. 3**. For pure Ti, the energy barrier decreases from 115.7 eV/atom at 0 GPa to 88.1 eV/atom at 15 GPa, showing that increasing pressure reduces the barrier and facilitates the transformation. Among the solutes, Al and Mo consistently result in the lowest energy barriers, with Al decreasing from 90.3 eV/atom at 0 GPa to 71.4 eV/atom at 15 GPa, and Mo decreasing from 93.1 eV/atom at 0 GPa to 74.6 eV/atom at 15 GPa. In contrast, Sn and Zr exhibit higher barriers, with Zr having the highest barrier at 0 GPa (102.3 eV/atom) and Sn at 15 GPa (81.3 eV/atom). Notably, all solutes reduce the energy barrier compared to pure Ti, with Al being the most effective in lowering the barrier across all pressures. These trends highlight the significant influence of solute atoms and pressure on the phase transformation, with Al and Mo being the most effective in lowering the barrier, while Sn and Zr also contribute to reducing the barrier compared to Pure Ti.

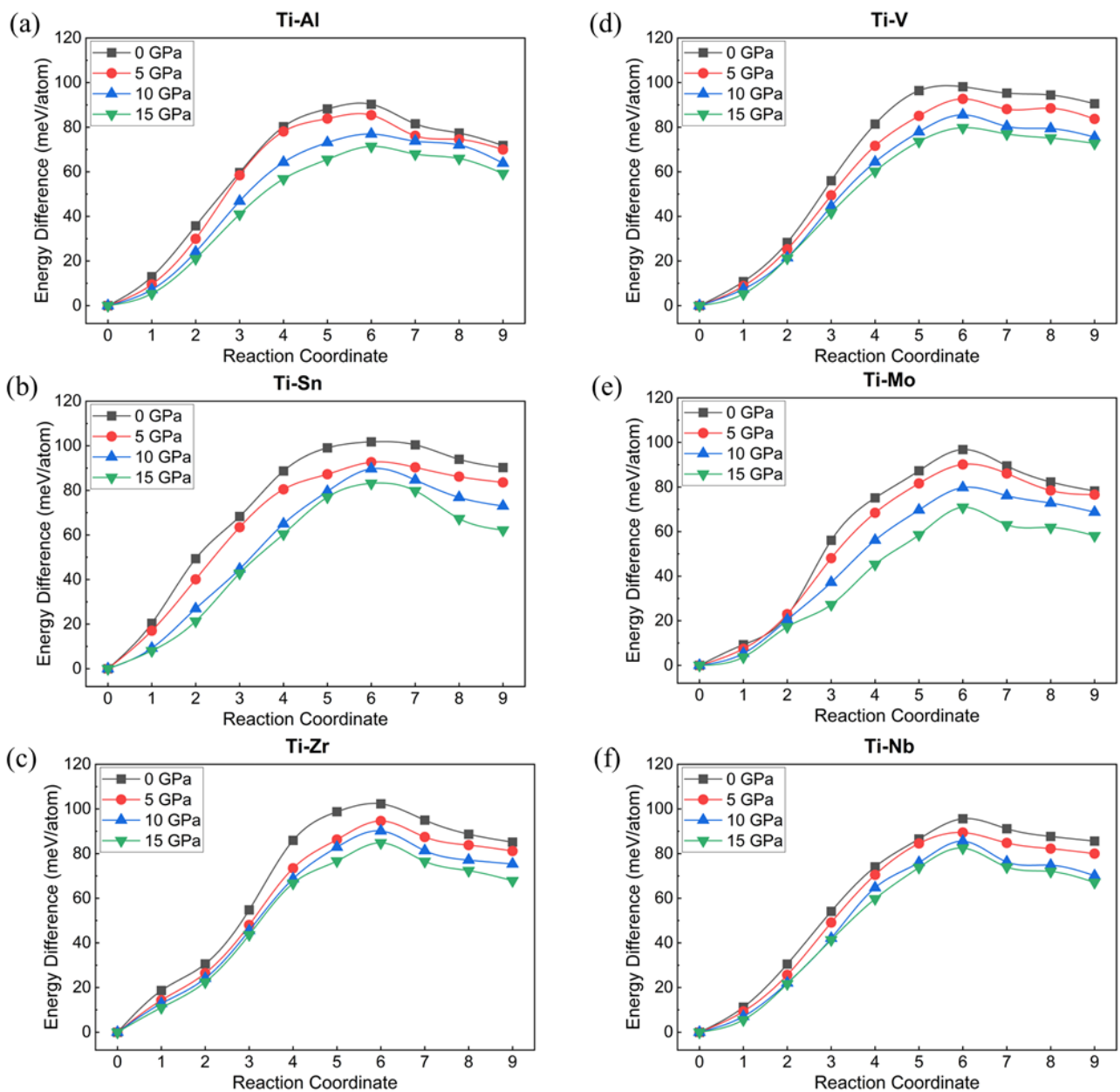

**Fig. 2.** Enthalpy landscape of the α-β phase transformation with a specific solute atom introduced along the transformation path illustrated in **Fig. 1**(a), calculated at pressures of 0 GPa, 5 GPa, 10 GPa, and 15 GPa. The energy profiles reveal how pressure influences the thermodynamic stability and transformation barrier between the two phases in the presence of the solute, offering insight into pressure-assisted phase transitions in alloyed titanium systems.

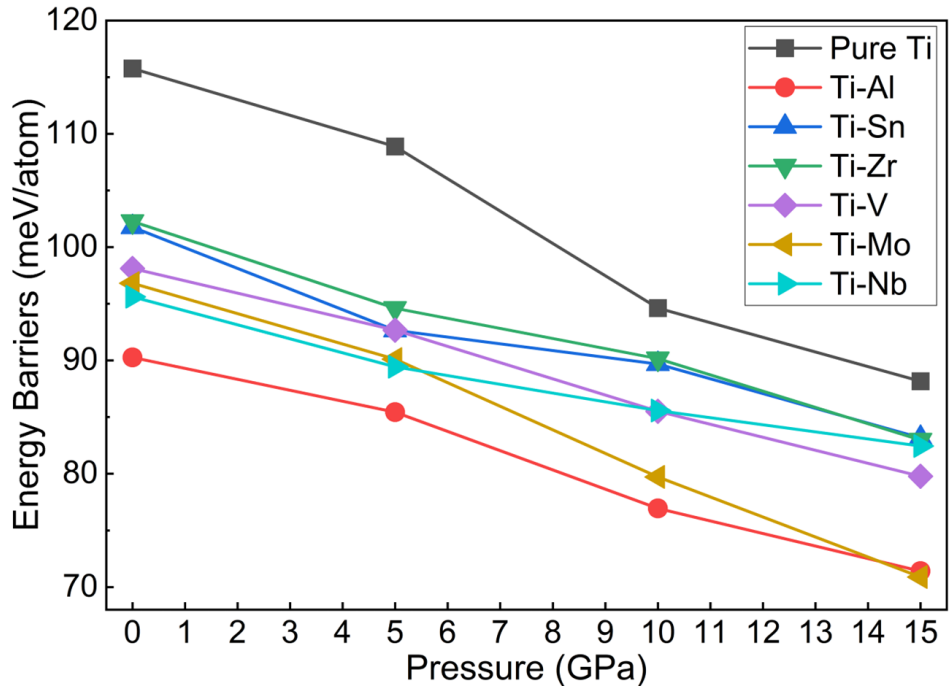

**Fig. 3.** Dependence of energy barrier as a function of pressure for different alloy systems. The energy barrier (activation enthalpy) is defined as the enthalpy difference between the transition state and the initial hcp phase from **Fig. 1**(b) and **Fig. *2***(a-f).

3.2 Electron charge density difference analysis

To reveal the physical origin of the solute effect phase stability, the ECDD and BCA were calculated for all substitutional solutes in the initial α phase and final β phase. The results are shown in **Fig. 4** and **Fig. 5**, respectively. In case of α phase (**Fig. 4**), ECDD values of Al decrease from 0.476 eV at 0 GPa to 0.332 eV at 15 GPa, indicating strong but gradually weakening Al-Ti interactions. This behavior aligns with role of Al as an α-phase stabilizer, enhancing thermodynamic stability through consistent charge transfer. In contrast, Sn shows a significant decrease in charge density, from 0.673 eV at 0 GPa to 0.255 eV at 15 GPa, reflecting a reduction in electron accumulation around the solute atom. This trend suggests that the influence of Sn on the local electronic structure diminishes with increasing pressure, potentially destabilizing the α-phase. For V, charge density values remain high, decreasing slightly from 0.693 eV at 0 GPa to 0.532 eV at 15 GPa, reflecting strong electron accumulation and the local bonding stability with varying pressure. Additionally, Mo also shows significant electron accumulation, decreasing from 0.819 eV at 0 GPa to 0.532 eV at 15 GPa, indicating strong bonding interactions that stabilize the α-phase. In comparison, Nb exhibits a more moderate influence, with values decreasing from 0.411 eV at 0 GPa to 0.114 eV at 15 GPa, suggesting a persistent but less pronounced effect on the electronic structure. Differently, Zr exhibits consistent electron depletion, with values decreasing from -0.004 eV at 0 GPa to -0.486 eV at 15 GPa, indicating weakened local bonding and a tendency to destabilize the α-phase.

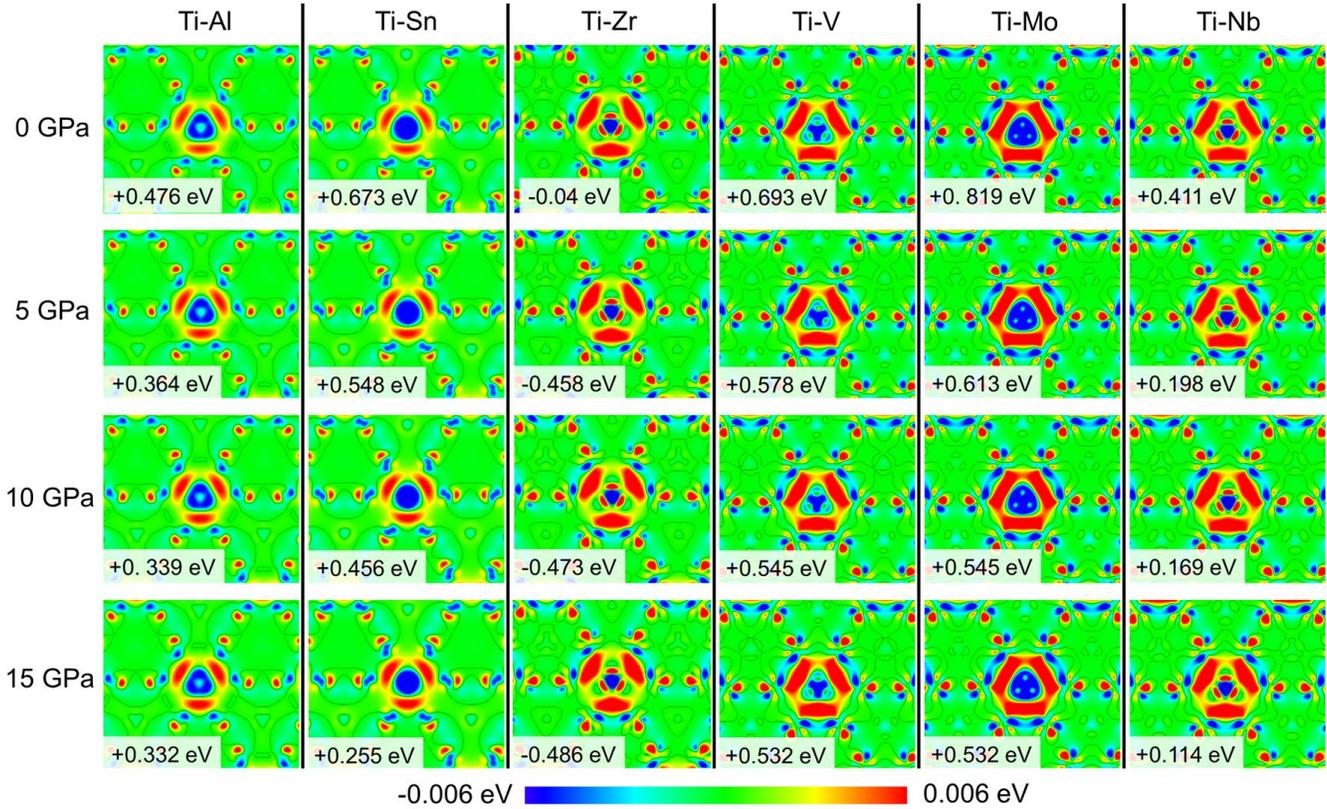

**Fig. 4**. Differential electron charge density maps in the (0001) plane of α-phase systems under pressures ranging from 0 to 15 GPa. The visualizations highlight charge redistribution between solute and matrix atoms, with blue and red regions indicating electron depletion and accumulation, respectively. Bader charge analysis was employed to quantify the charge transfer, and the corresponding values are annotated in each subfigure.

For β phase, as shown in **Fig. 5**, the charge density transitions from positive values (0.524 eV at 0 GPa) to negative values (-0.043 eV at 15 GPa), indicating a shift from electron accumulation to depletion with increasing pressure. This suggests that Al may destabilize the bcc phase at higher pressures. In contrast, Sn maintains positive charge density values, albeit decreasing from 0.336 eV at 0 GPa to 0.194 eV at 15 GPa, reflecting a persistent but weakening stabilizing influence. Zr exhibits consistent electron depletion across all pressures, with values decreasing from -0.312 eV at 0 GPa to -0.491 eV at 15 GPa, underscoring its destabilizing role in the bcc structure. For V, the charge density remains negative, ranging from -0.105 eV at 0 GPa to -0.013 eV at 15 GPa, indicating electron depletion and a lack of stabilizing contribution. Mo and Nb show significant electron accumulation at low pressures (0.677 eV and 0.495 eV at 0 GPa, respectively), which decreases to 0.043 eV and 0.211 eV at 15 GPa, suggesting a stabilizing effect that diminishes with increasing pressure.

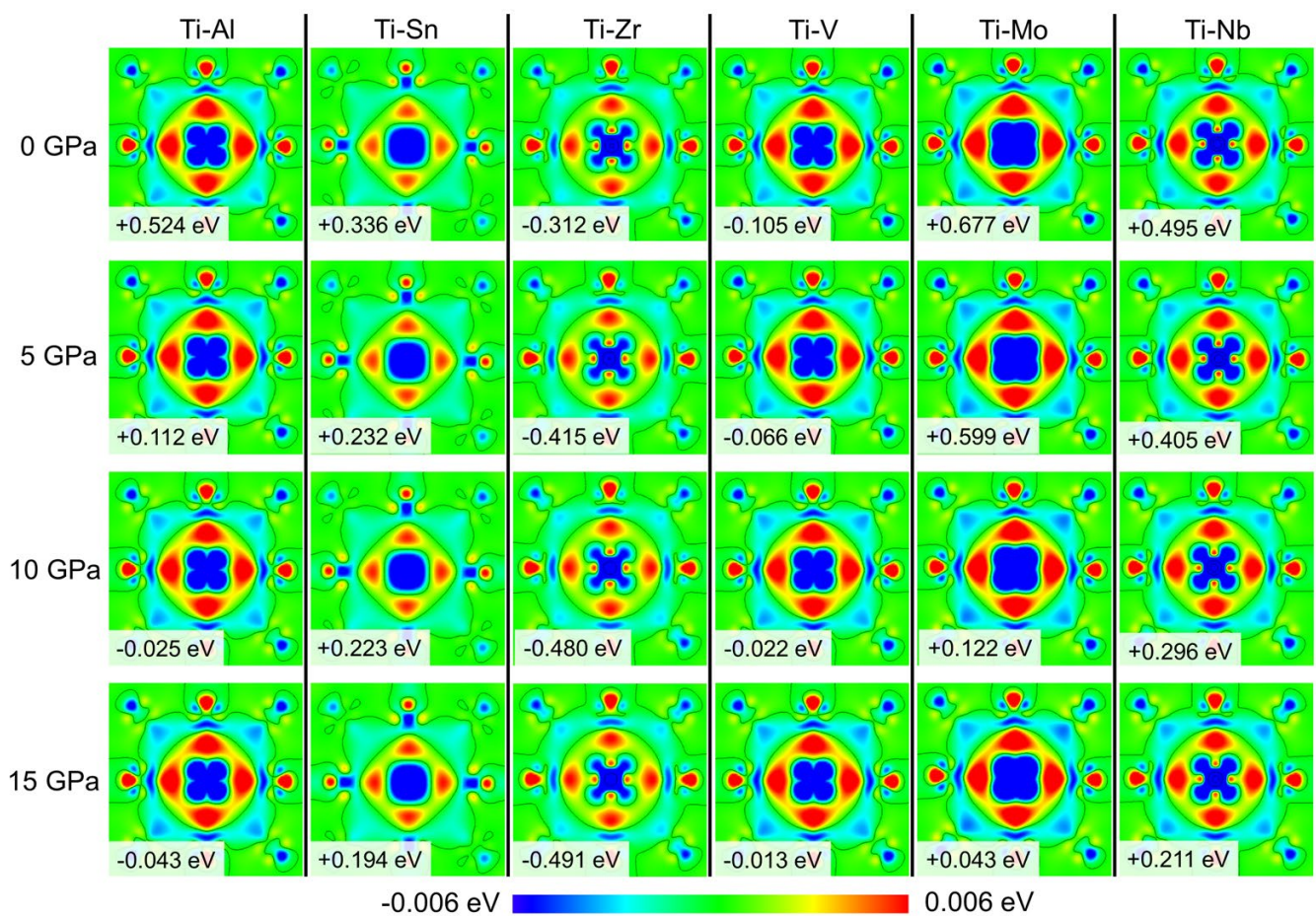

**Fig. 5.** Differential electron charge density distributions in the (001) plane of β-phase systems subjected to pressures from 0 to 15 GPa. The maps illustrate the electron redistribution induced by the interaction between solute and matrix atoms. Regions of electron depletion and accumulation are represented by blue and red areas, respectively. Quantitative analysis of charge transfer was conducted using the Bader method, with the resulting values indicated in each panel.

3.3 The density of states (DOS) analysis

**Fig. 6** presents the total density of states (TDOS) for the initial α-phase of Ti-based alloys under varying pressures (0–15 GPa), with the Fermi energy ($E_\text{F}$) set at 0 eV. The TDOS curves exhibit a systematic shift away from the Fermi level as pressure increases, indicating a weakening of bonding interactions within the α-phase. At 0 GPa, the TDOS curve shows significant electronic states near the Fermi level, reflecting strong bonding characteristics typical of the α-phase. As pressure increases to 5 GPa, the curve shifts slightly away from $E_F$, suggesting a reduction in bonding strength. This trend continues at 10 GPa and 15 GPa, where the TDOS curves are further displaced from the Fermi level, indicating a progressive weakening of bonding interactions under higher pressures. The shift in TDOS curves is consistent with the pressure-induced destabilization of the α-phase, as the electronic states near the Fermi level play a critical role in determining bonding stability. The observed behavior aligns with the known pressure-driven phase transformation from hcp (α-phase) to bcc (β-phase) in Ti-based alloys, where increased pressure reduces the stability of the α-phase.

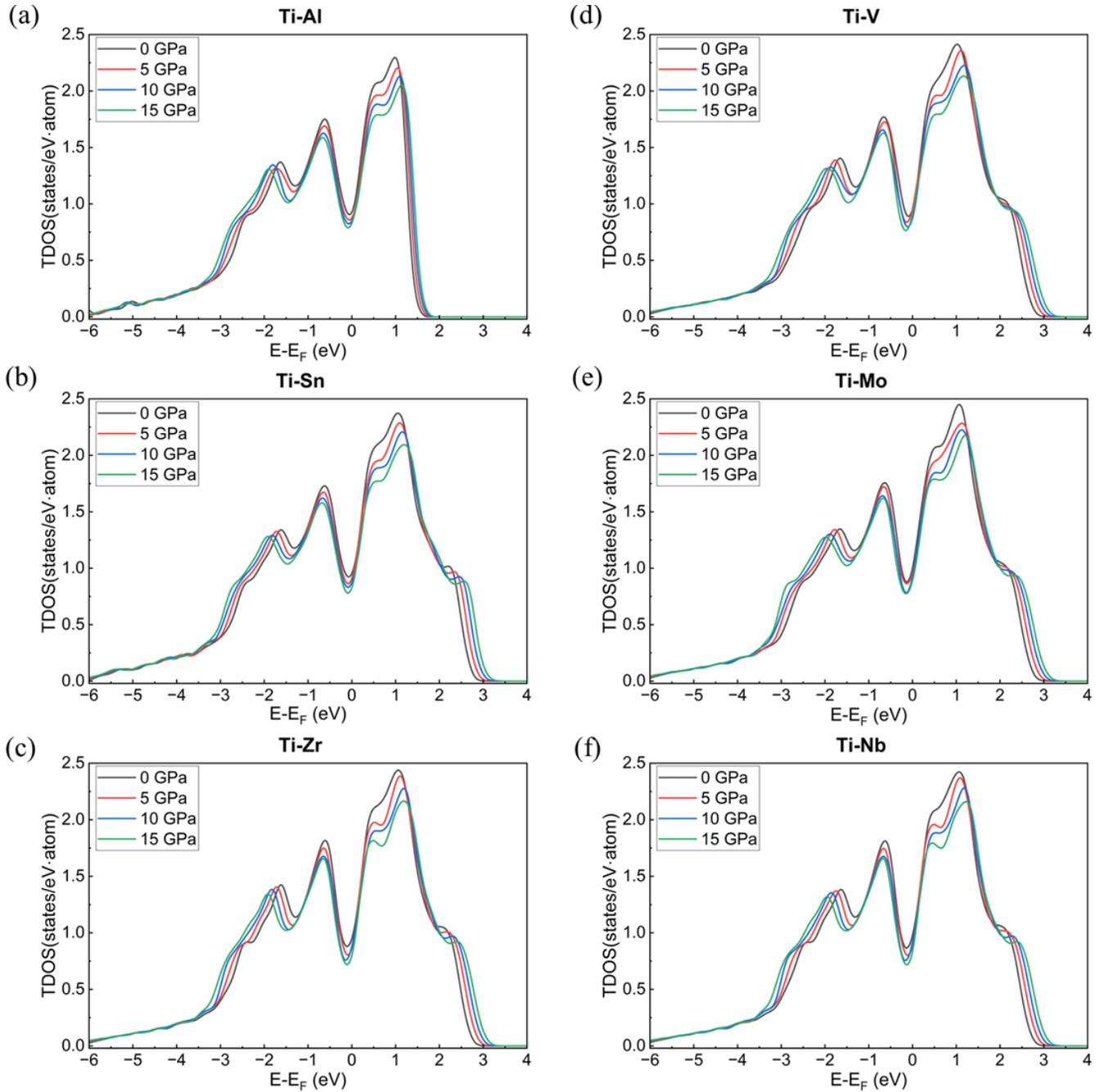

**Fig. 6.** Total density of states (TDOS) of the initial α-phase structure under varying pressure conditions from 0 to 15 GPa. The Fermi energy is set at 0 eV for reference in all plots. With increasing pressure, the TDOS curves progressively shift away from the Fermi level, reflecting a redistribution of electronic states.

A similar trend is observed in the bcc structure (**Fig. 8**). At 0 GPa, the TDOS curve shows significant electronic states near the Fermi level, characteristic of strong bonding in the β-phase. As pressure increases to 5 GPa, the curve shifts slightly away from $E_F$, suggesting a reduction in bonding strength. This trend continues at 10 GPa and 15 GPa, where the TDOS curves are further displaced from the Fermi level, reflecting a progressive weakening of bonding interactions under higher pressures. However, the shift in TDOS curves is smaller in the β-phase compared to the α-phase, indicating that the β-phase retains stronger bonding characteristics under pressure. This difference aligns with the known pressure-driven phase transformations in Ti-based alloys, where the β-phase exhibits greater stability under compression relative to the α-phase. The smaller shift in the β-phase suggests that its electronic structure is less sensitive to pressure-induced changes, which may contribute to its enhanced stability under high-pressure conditions.

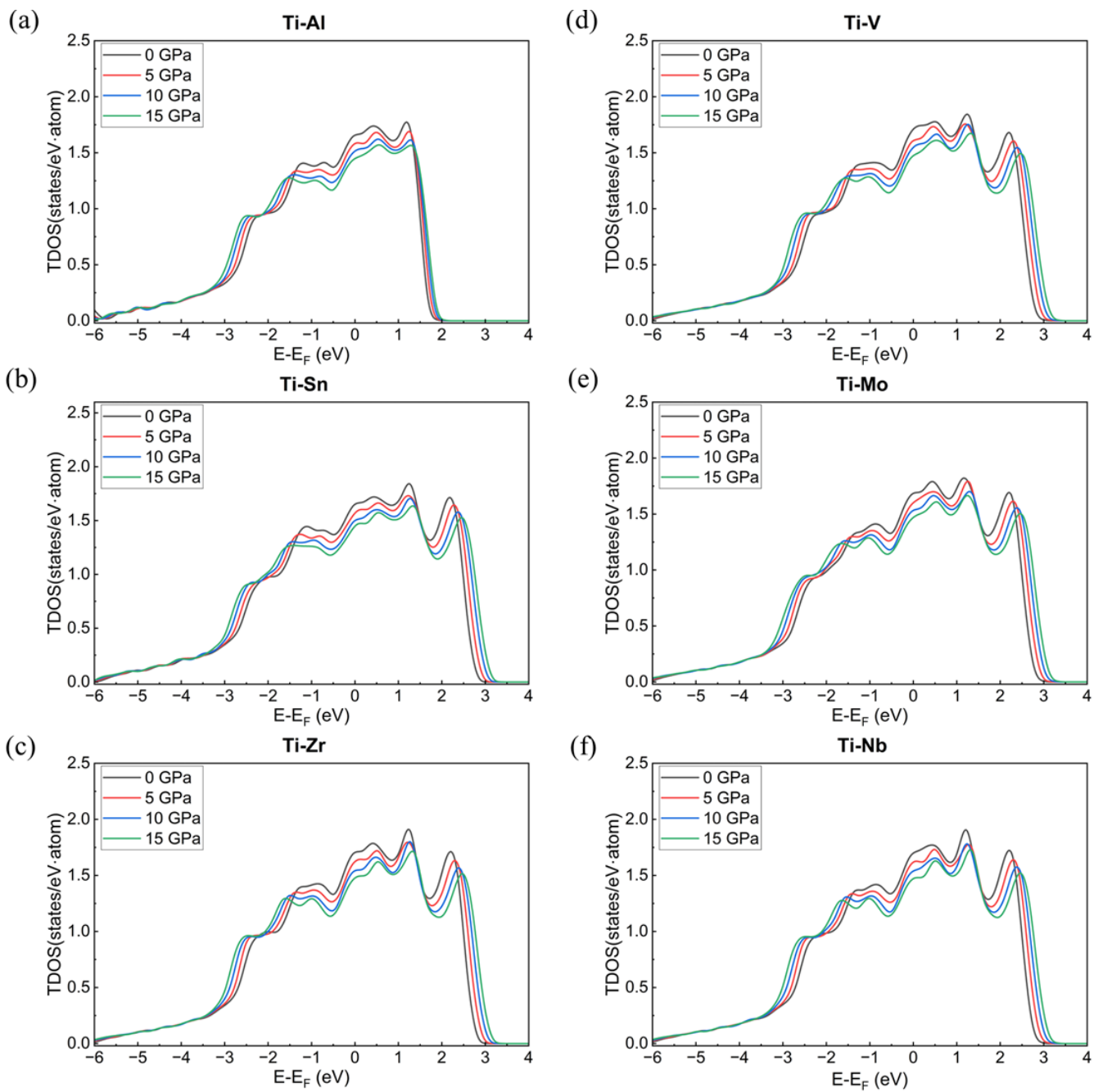

**Fig. 7.** Total density of states (TDOS) of the initial β-phase structure computed under varying pressure conditions from 0 to 15 GPa. In all the plots, the Fermi energy is aligned at 0 eV to provide a consistent reference point. As the applied pressure increases, the TDOS curves exhibit a noticeable shift away from the Fermi level, indicating a redistribution of electronic states.

To gain insights into the nature of chemical bonding under varying pressures, the electron charge density difference (ECDD) was calculated by comparing systems with and without applied pressure. As shown in **Fig. 9** and **Fig. 10**, the β phase exhibits a greater tendency for electron redistribution, indicating a loss of stability with increasing pressure, particularly in the Ti-Sn and Ti-Mo systems.

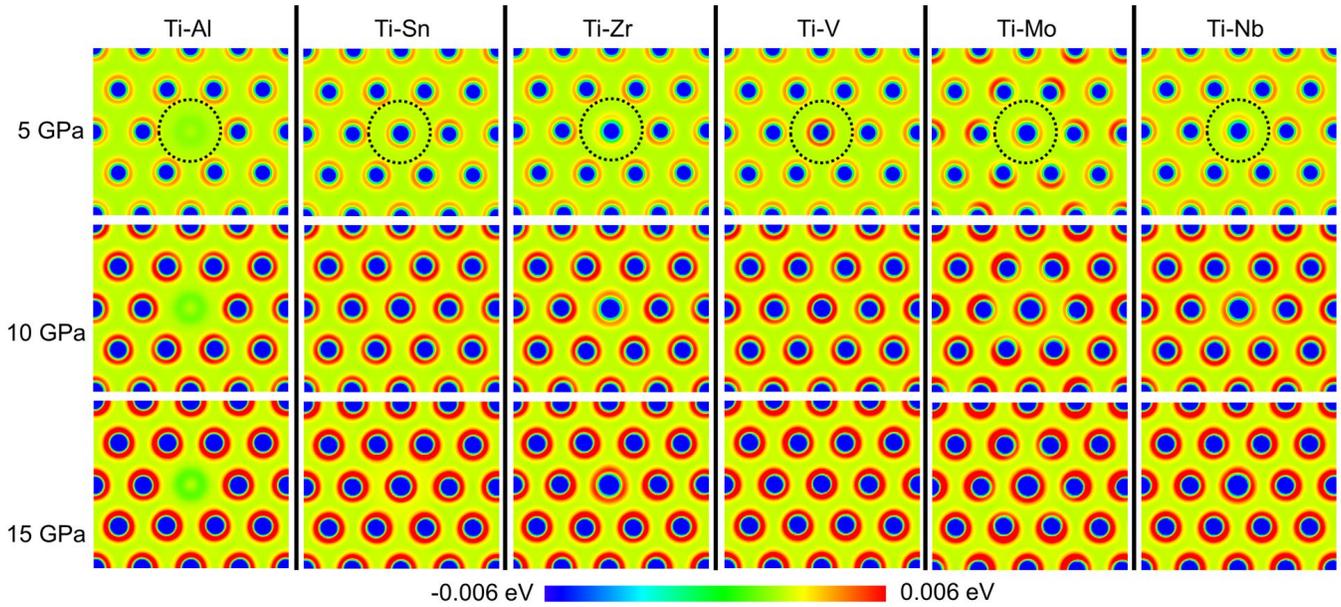

**Fig. 8**. Differential electron charge density maps comparing the compressed and pressure-free states of the α-phase structure. These maps highlight the redistribution of electronic charge resulting from the application of external pressure. Blue regions represent electron depletion, while red regions indicate electron accumulation. Solute atoms are marked with dashed outlines to highlight their positions within the lattice for each alloy system at a pressure of 5 GPa.

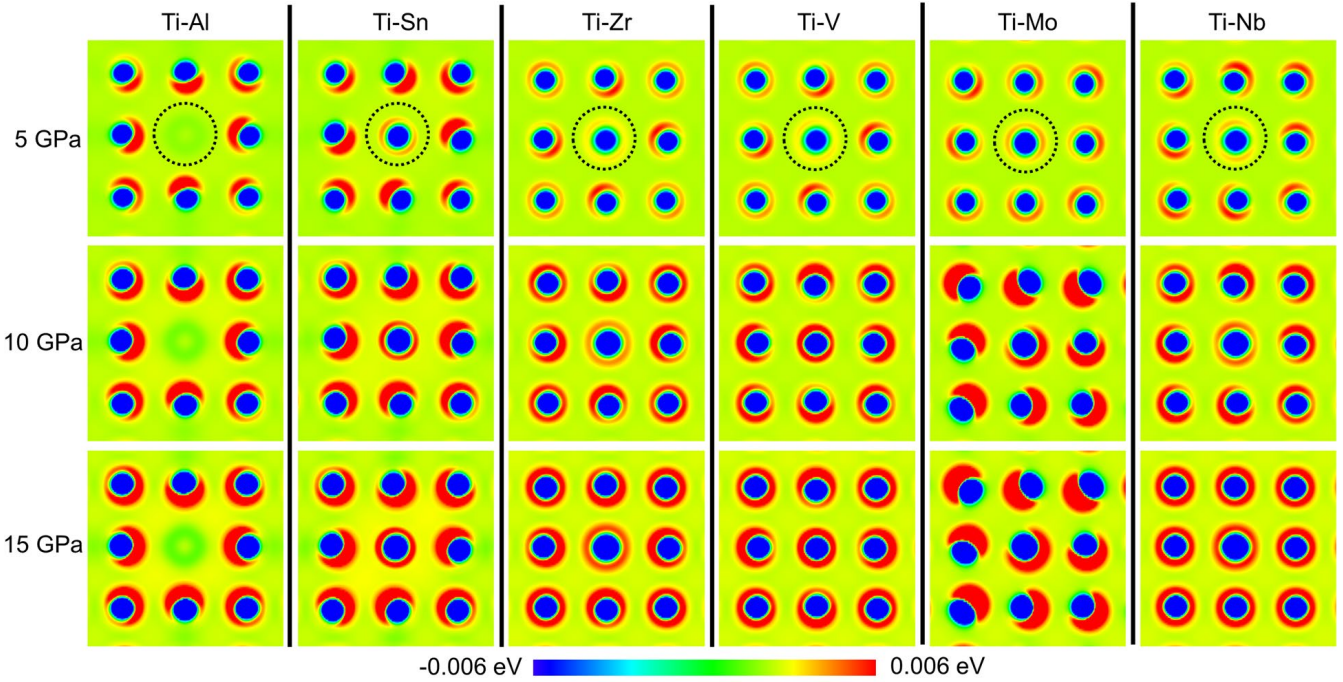

**Fig. 9.** Differential electron charge density maps illustrating the electronic response of the β-phase structure to applied pressure, obtained by comparing the compressed and pressure-free states. The visualizations capture the redistribution of electron density induced by compression. Blue regions indicate electron depletion, while red regions denote electron accumulation. Solute atoms are marked with dashed outlines to highlight their positions within the lattice for each alloy system at a pressure of 5 GPa.

**4 Discussion**

Our calculations reveal that phase transformations in Ti alloys can be effectively modified by adding solute and applying external mechanical loading. These findings demonstrate that the energy barrier for phase transitions is highly sensitive to both compositional changes and external stresses, providing valuable insights into the tunability of phase stability. Specifically, the incorporation of solute elements reduces the intrinsic energy barrier for the transformation, while external loading further lowers the barrier by altering the mechanical energy landscape. These synergistic effects highlight the potential for precise control of phase transformations in Ti alloys, especially under high-temperature service conditions.

To elucidate the mechanisms influencing phase transformations, we reference the decomposition of the chemical Gibbs free energy ($\Delta G^{chem}$) [40] expressed as the following:

$$\Delta G^{chem} = \Delta G^{SD} + \Delta G^{Str} + \Delta G^{CT} + \Delta G^{mech} \tag{1}$$

Here, $\Delta G^{SD}$ is attributed to solute drag, which arises from the interaction of solute atoms with the migrating interfaces during phase transformation. This term reflects the resistance provided by solute atoms to the transformation process. $\Delta G^{Str}$ accounts for the energy required to overcome the structural rearrangement between the hcp and bcc lattices. $\Delta G^{CT}$ represents the interface curvature energy, which depends on the shape and morphology of the transforming phases and contributes to the overall transformation energetics. Finally, $\Delta G^{mech}$ represents the mechanical energy contribution, which is influenced by external loading and strain energy. Together, these terms provide a comprehensive framework for understanding and manipulating phase transformations.

As shown in **Fig. 10**, the schematic illustration provided further clarifies how solute addition, external loading, and/or temperature synergistically reduce the energy barrier for phase transformation. In the static state, the energy barrier for the α-β transition is defined by $\Delta G^{S}_{\alpha \to \beta}$, representing the intrinsic energy required under equilibrium conditions. This barrier, though significant, can be modified through external interventions. Under dynamic conditions, the application of external loading contributes to $\Delta G^{mech}$, introducing strain energy that facilitates the phase transformation. This effect reduces the energy barrier to $\Delta G^{D}_{\alpha \to \beta}$, as shown in the second panel of the schematic. This reduction reveals the role of mechanical forces in driving phase transitions, a concept particularly relevant for deformation-assisted transformations. The addition of solute elements further decreases the energy barrier, as represented by $\Delta G^{D+So}_{\alpha \to \beta}$. Solute atoms interact with the interfaces, reducing $\Delta G^{SD}$ by altering the drag forces that resist transformation. Additionally, solutes may influence by modifying the curvature energy $\Delta G^{CT}$ at the interface, thereby enhancing the transformation kinetics. At elevated temperatures, the energy barrier is further diminished, resulting in $\Delta G^{D+So+T}_{\alpha \to \beta}$. Temperature facilitates the overcoming of structural rearrangement and amplifies thermal activation effects, further accelerating phase transitions. The combination of these factors demonstrates the importance of high-temperature conditions in reducing the energy barriers for phase transformation.

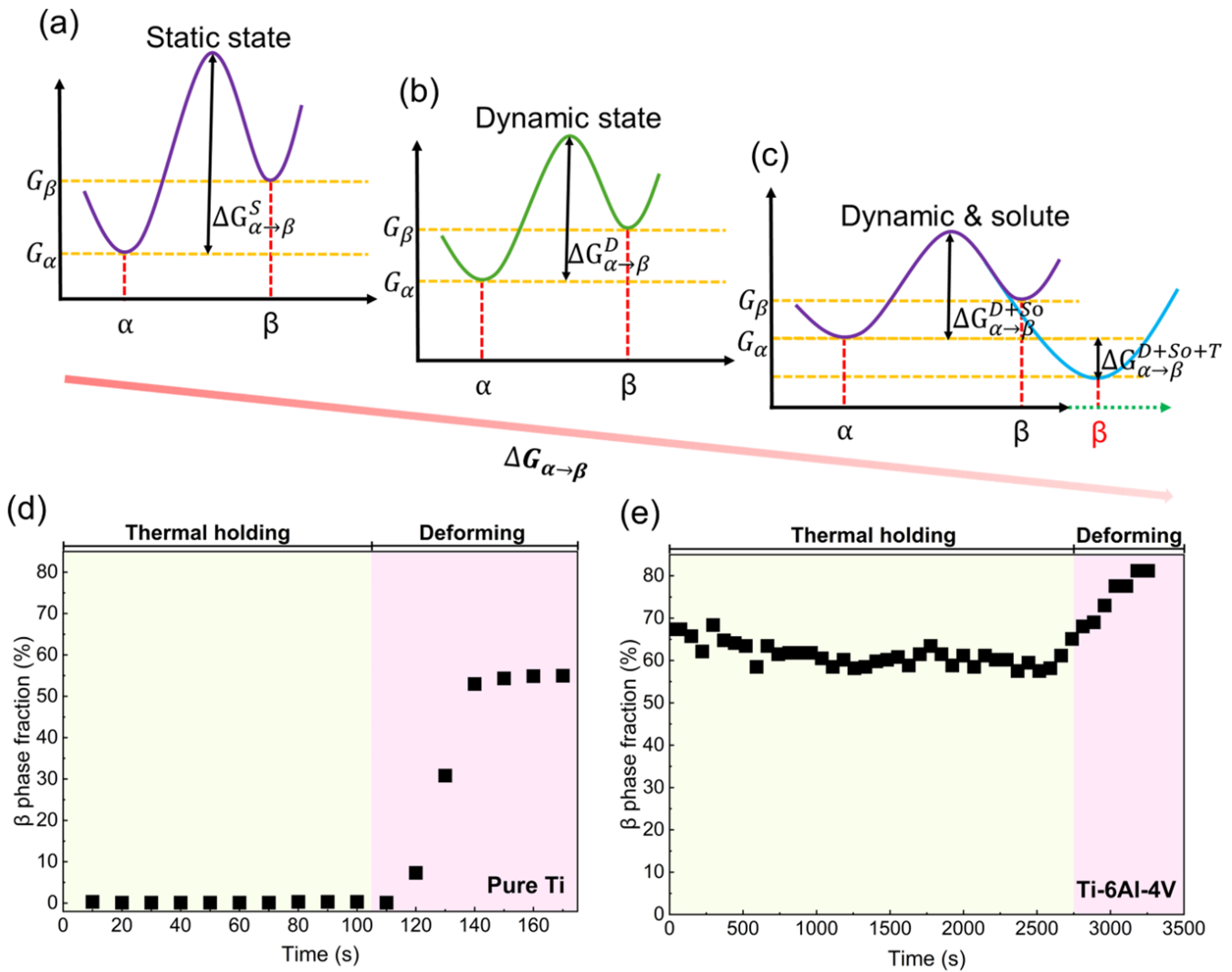

**Fig. 10.** Schematic illustrations of Gibbs free energy variations between the α and β phases of titanium under different conditions. (a) Gibbs free energy profiles of pure Ti in a static, equilibrium state without external loading. (b) Free energy evolution under dynamic loading conditions, highlighting the effect of mechanical stress on phase stability in pure Ti. (c) Similar to (b) but incorporating the effect of alloying at a target temperature *T*, illustrating how solute additions modify the energy landscape and promote phase transformations. (d) Experimentally observed β-phase fraction changes in pure Ti during thermal holding and mechanical deformation at 860 °C. (e) Corresponding β-phase fraction evolution in the Ti-6Al-4V alloy under the same sequence of processes at 930 °C.

The integration of solute engineering, external loading, and high-temperature conditions provides a systematic pathway for controlling phase transformations in Ti alloys. Based on our findings, we propose a design map (**Fig. 11**) that encompasses the effects of solute elements and external pressure, offering a comprehensive tool for guiding the development of new Ti alloys. The contour plot highlights distinct zones where specific combinations of solute elements and pressure produce optimal conditions for phase control. These zones provide a visual guide for selecting solute elements and processing pressures to tailor phase stability and transformation kinetics.

For instance, the contour map reveals that the addition of Al or Mo results in the lowest energy barrier at 0 GPa (ambient pressure), as indicated by the light blue region. This observation suggests that these solutes inherently improve the phase transition by lowering the energy barrier, making them particularly favorable for facilitating phase control in titanium alloys under no external pressure. When external pressure is applied, the energy barrier is further reduced, as seen in the downward progression of the contour lines towards darker blue with increasing pressure along the x-axis. This indicates that combining Al or Mo solutes with applied pressure enhances the phase transition process even further. For other alloy systems, regions on the map with moderate energy barriers (e.g., green zones) demonstrate that applying suitable external loading can shift these systems into lower energy states (blue regions). This highlights the potential of pressure-assisted phase control to optimize phase transformations in alloys with less favorable intrinsic solute effects.

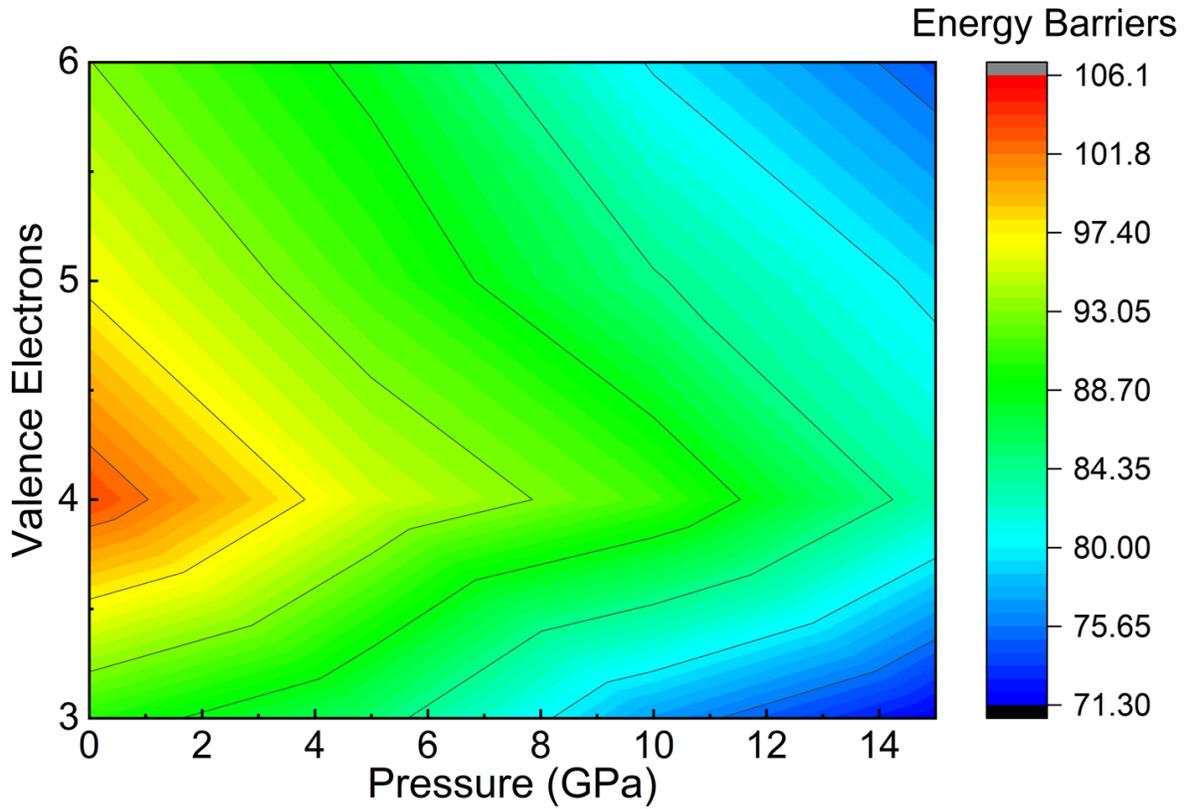

**Fig. 11.** Design map for phase control in titanium-based systems through compositional tailoring and/or mechanical stress modulation. The schematic illustrates how the introduction of specific alloying elements and the application of external stress can be strategically used to stabilize or transform between α and β phases.

To validate the reliability of the current simulation results, we incorporate relevant experimental observations as shown in **Fig. 10** (d, e), which presents the evolution of the β phase fraction in pure Ti and Ti-Al6-V4 during both the holding and deformation stages, as captured by in-situ neutron diffraction [35]. The β phase fraction reaches equilibrium during the heating process. However, the α-β phase transformation resumes upon the application of external stress, as observed in the deformed region of **Fig. 11**. This result indicates that mechanical loading facilitates the α-β phase transition by lowering the transformation barrier, a phenomenon further supported by our DFT calculations. Moreover, recent experimental studies have shown that the addition of solute elements, such as Mo, can reduce the transformation barrier by altering the intermediate structures involved in the transition [14]. Therefore, the combined effects of alloying and mechanical loading provide an effective strategy for controlling the final phase fractions in titanium, thereby enabling the tailoring of its properties.

## 5 Conclusions

In this study, we have investigated the effects of solute elements and external pressure on the energy barriers associated with α-β phase transformations in Ti alloys via using DFT combined with generalized solid-state NEB calculations. The key findings are summarized as follows:

(1) Both solute addition and hydrostatic pressure reduce the activation enthalpy of the α-to-β phase transition. For pure Ti, the energy barrier decreases from 115.7 meV/atom at 0 GPa to 88.1 meV/atom at 15 GPa, promoting phase transformation with increasing pressure. Solute additions further lower the energy barrier, with adding Al and Mo more effective than Sn, Zn, V and V. For instance, Al reduces the energy barrier from 90.3 meV/atom at 0 GPa to 71.4 meV/atom at 15 GPa, while Mo reduces it from 93.1 meV/atom to 74.6 meV/atom over the same pressure range. In contrast, adding Sn or Zr exhibits relatively higher energy barriers, for example, 81.3 meV/atom at 15 GPa for adding Sn to Ti. These results indicate that solute effects are solute-specific and pressure-dependent.

(2) Electron charge density difference (ECDD) and Bader charge analysis (BCA) indicate that solute elements such as Mo, V, and Nb enhance β-phase stability by increasing electron accumulation around the solute atoms, especially under moderate pressures (5-15 GPa). For example, the charge density around Mo in the β phase remains positive and across pressures, ranging from 0.677 eV at 0 GPa to 0.043 eV at 15 GPa, suggesting robust Ti–Mo bonding. These electronic effects help explain the improved phase stability and lower transformation barriers in solute-stabilized Ti alloys.

(3) The combined effect of solute alloying and external pressure synergistically reduces the energy barrier of α–β transformation more effectively than either factor alone. The design map highlights Al and Mo as optimal solutes, with pressure further enhancing their impact on the phase transformation. The DOS analysis and experimental comparisons confirm the practical relevance of this phase-control strategy.

**Declaration of Competing Interest**

The authors declare that they have no known competing financial interests or personal relationships that could have appeared to influence the work reported in this paper.

**Acknowledgement**

The authors acknowledge the Digital Research Alliance of Canada for providing computing resources. Y.Z. acknowledges the financial support from NSERC, Discovery Grant #RGPIN-2018-05731, Y.Z.

acknowledge the financial support from the Centre for Analytics and Artificial Intelligence Engineering (CARTE) Seed Funding program, Data Sciences Institute Catalyst Grant, and NSERC Alliance Grants—Missions ALLRP 570708-2021. This research is part of the University of Toronto's Acceleration Consortium, which receives funding from the Canada First Research Excellence Fund (CFREF).

# References

[1] Y.-W. Cui, L. Wang, L.-C. Zhang, Towards load-bearing biomedical titanium-based alloys: From essential requirements to future developments, Progress in Materials Science 144 (2024) 101277.
[2] G. Xu, X. Zhao, W. Xia, Q. Yue, Z. Zheng, Y. Gu, Z. Zhang, A review on microstructure design, processing, and strengthening mechanism of high-strength titanium alloys, Progress in Natural Science: Materials International (2025).
[3] T. Song, Z. Chen, X. Cui, S. Lu, H. Chen, H. Wang, T. Dong, B. Qin, K.C. Chan, M. Brandt, X. Liao, S.P. Ringer, M. Qian, Strong and ductile titanium–oxygen–iron alloys by additive manufacturing, Nature 618(7963) (2023) 63-68.
[4] S. Wei, K.-S. Kim, J. Foltz, C.C. Tasan, Discovering Pyramidal Treasures: Multi-Scale Design of High Strength–Ductility Titanium Alloys, Advanced materials 36(33) (2024) 2406382.
[5] B. Guo, S.L. Semiatin, J.J. Jonas, Dynamic transformation during the high temperature deformation of two-phase titanium alloys, Materials Science and Engineering: A 761 (2019) 138047.
[6] J. Koike, Y. Shimoyama, I. Ohnuma, T. Okamura, R. Kainuma, K. Ishida, K. Maruyama, Stress-induced phase transformation during superplastic deformation in two-phase Ti–Al–Fe alloy, Acta Materialia 48(9) (2000) 2059-2069.
[7] J. Bustillos, A. Wakai, K. Singh, C. Tian, A. Dass, S.P. Akula, A. Das, A. Moridi, Alloy amalgamation: Unlocking the co-existence of multiple phases in additively manufactured titanium alloys, Applied Materials Today 42 (2025) 102606.
[8] Y. Tian, L. Zhang, D. Wu, R. Xue, Z. Deng, T. Zhang, L. Liu, Achieving stable ultra-low elastic modulus in near-β titanium alloys through cold rolling and pre-strain, Acta Materialia 286 (2025) 120726.
[9] S. Sun, H. Fang, J. Hao, B. Zhu, X. Ding, R. Chen, Formation behavior of subcrystals and its strengthening and toughening mechanism by coupling with α phase in titanium alloys during forging at various temperatures, Journal of Materials Processing Technology 336 (2025) 118705.
[10] Z. Zheng, H. Wu, S. Zhang, Z. Liao, S. Wu, F. Cheng, The microstructure evolution and embrittlement mechanism in the heat-affected zone of thick-plate titanium alloys fabricated by gas metal arc welding, Journal of Materials Processing Technology 335 (2025) 118657.
[11] C. Ghosh, C. Aranas, J.J. Jonas, Dynamic transformation of deformed austenite at temperatures above the Ae3, Progress in Materials Science 82 (2016) 151-233.
[12] H. Shahmir, M. Nili-Ahmadabadi, H.S. Kim, T.G. Langdon, Significance of adiabatic heating on phase transformation in titanium-based alloys during severe plastic deformation, Materials Characterization 203 (2023) 113091.
[13] K. Wang, B. Qu, J. Zhao, B. He, G. Liu, Accelerated stress relaxation with simultaneously enhanced strength of titanium alloy by phase transformation and stress-induced twinning, Scripta Materialia 238 (2024) 115761.
[14] X. Fu, X.-D. Wang, B. Zhao, Q. Zhang, S. Sun, J.-J. Wang, W. Zhang, L. Gu, Y. Zhang, W.-Z. Zhang, W. Wen, Z. Zhang, L.-q. Chen, Q. Yu, E. Ma, Atomic-scale observation of non-classical nucleation-mediated phase transformation in a titanium alloy, Nature Materials 21(3) (2022) 290-296.

[15] J. Ying, S. Xu, G. Liu, S. Pan, Q. Fan, X. Cheng, Enhanced dynamic compression properties of a low density near-α titanium alloy associated with deformation induced laminated microstructure and dynamic segregation, Acta Materialia 285 (2025) 120653.
[16] S. Zhang, Y. Zhu, F. Zhang, X. Guo, Y. Xu, H. Wang, Y. Yin, H. Liu, Z. Wei, Z. Liao, W. Hu, Y. Lv, L. Chen, S. Li, Novel flat-top laser-aided cold metal transfer additive manufacturing for titanium alloy: Arc characteristics, microstructure, and tensile properties, Journal of Materials Processing Technology 327 (2024) 118379.
[17] F. Brumbauer, N.L. Okamoto, T. Ichitsubo, W. Sprengel, M. Luckabauer, Minor additions of Sn suppress the omega phase formation in beta titanium alloys, Acta Materialia 262 (2024) 119466.
[18] Y. Jia, H. Su, S. Cao, R. Shi, Y. Ma, Q. Wang, S. Huang, R. Zhang, Q. Hu, Y. Zheng, S. Zheng, J. Lei, R. Yang, Fabrication of highly heterogeneous precipitate microstructure in an α/β titanium alloy, Acta Materialia 279 (2024) 120302.
[19] W. Zhang, X. Qi, S. Zhong, K. Wang, S. Zhang, Y. Jiao, A. Li, H. Xu, J. Chen, G. Fang, W. Liu, Laser powder bed fusion of metastable β titanium alloys: Enhanced strength and plasticity through simultaneous activation of multiple deformation mechanisms, Additive Manufacturing 93 (2024) 104437.
[20] K. Li, D. Zhao, Y. Xing, W. Chen, J. Zhang, J. Sun, Kink-mediated high strength and large ductility via nanocrystallization in a tough titanium alloy, Acta Materialia 273 (2024) 119963.
[21] S. Balachandran, S. Kumar, D. Banerjee, On recrystallization of the α and β phases in titanium alloys, Acta Materialia 131 (2017) 423-434.
[22] X. Pang, Z. Xiong, C. Yao, J. Sun, R.D.K. Misra, Z. Li, Strength and ductility optimization of laser additive manufactured metastable β titanium alloy by tuning α phase by post heat treatment, Materials Science and Engineering: A 831 (2022) 142265.
[23] E. Hoareau, T. Billot, C. Deleuze, S. Andrieu, E. Maawad, Y. Thebault, A. Pugliara, B. Viguier, M. Dehmas, Phase transformation kinetics in Ti 575 titanium alloy during heat treatment: Role of the initial microstructure during ageing, Journal of Alloys and Compounds 1009 (2024) 176906.
[24] J. Li, S. Liu, C. Xia, H. Xu, J. Cai, W. Xiao, X. Huang, C. Ma, Effects of pre-tension on the microstructural stability and mechanical properties of a near-alpha high temperature titanium alloy, Journal of Alloys and Compounds 1010 (2025) 177506.
[25] P. Pesode, S. Barve, A review—metastable β titanium alloy for biomedical applications, Journal of Engineering and Applied Science 70(1) (2023) 25.
[26] N.N. Cherenda, A.V. Basalai, V.I. Shymanski, V.V. Uglov, V.M. Astashynski, A.M. Kuzmitski, A.P. Laskovnev, G.E. Remnev, Modification of Ti-6Al-4V alloy element and phase composition by compression plasma flows impact, Surface and Coatings Technology 355 (2018) 148-154.
[27] J. Ballor, T. Li, F. Prima, C.J. Boehlert, A. Devaraj, A review of the metastable omega phase in beta titanium alloys: the phase transformation mechanisms and its effect on mechanical properties, International Materials Reviews 68(1) (2023) 26-45.
[28] Y. Huang, F. Zhang, Y. Xiong, T. Dai, Q. Wan, Selective laser melting processing of heterostructured Ti6Al4V/FeCoNiCrMo alloy with superior strength and ductility, Journal of Alloys and Compounds 978 (2024) 173435.
[29] E. Alabort, D. Barba, M.R. Shagiev, M.A. Murzinova, R.M. Galeyev, O.R. Valiakhmetov, A.F. Aletdinov, R.C. Reed, Alloys-by-design: Application to titanium alloys for optimal superplasticity, Acta Materialia 178 (2019) 275-287.
[30] L. Zhang, Y.-H. Li, Y.-Q. Gu, L.-C. Cai, Understanding controversies in the α-ω and ω-β phase transformations of zirconium from nonhydrostatic thermodynamics, Scientific Reports 9(1) (2019) 16889.
[31] J.C. Jamieson, Crystal Structures of Titanium, Zirconium, and Hafnium at High Pressures, Science 140(3562) (1963) 72-73.

[32] K. Binder, P. Virnau, Phase transitions and phase coexistence: equilibrium systems versus externally driven or active systems - Some perspectives, Soft Materials 19(3) (2021) 267-285.
[33] A. Rosenflanz, I.W. Chen, Kinetics of phase transformations in SiAlON ceramics: I. effects of cation size, composition and temperature, Journal of the European Ceramic Society 19(13) (1999) 2325-2335.
[34] B. Guo, W. Mao, Y. Chong, A. Shibata, S. Harjo, W. Gong, H. Chen, J.J. Jonas, N. Tsuji, Unexpected dynamic transformation from α phase to β phase in zirconium alloy revealed by in-situ neutron diffraction during high temperature deformation, Acta Materialia 242 (2023) 118427.
[35] B. Guo, H. Chen, Y. Chong, W. Mao, S. Harjo, W. Gong, Z. Zhang, J.J. Jonas, N. Tsuji, Direct observations of dynamic and reverse transformation of Ti-6Al-4V alloy and pure titanium, Acta Materialia (2024) 119780.
[36] D. Doraiswamy, S. Ankem, The effect of grain size and stability on ambient temperature tensile and creep deformation in metastable beta titanium alloys, Acta Materialia 51(6) (2003) 1607-1619.
[37] U. Dahmen, Orientation relationships in precipitation systems, Acta Metallurgica 30(1) (1982) 63-73.
[38] C. Cayron, Angular distortive matrices of phase transitions in the fcc–bcc–hcp system, Acta Materialia 111 (2016) 417-441.
[39] C. Cayron, Continuous atomic displacements and lattice distortion during fcc–bcc martensitic transformation, Acta Materialia 96 (2015) 189-202.
[40] I.-E. Benrabah, Y. Brechet, G. Purdy, C. Hutchinson, H. Zurob, On the origin of the barrier in the bainite phase transformation, Scripta Materialia 223 (2023) 115076.
[41] S.-j. Lv, S.-z. Wang, G.-l. Wu, J.-h. Gao, X.-s. Yang, H.-h. Wu, X.-p. Mao, Application of phase-field modeling in solid-state phase transformation of steels, Journal of Iron and Steel Research International 29(6) (2022) 867-880.
[42] J. Zhang, X. Li, D. Xu, C. Teng, H. Wang, L. Yang, H. Ju, H. Xu, Z. Meng, Y. Ma, Y. Wang, R. Yang, Phase field simulation of the stress-induced α microstructure in Ti–6Al–4 V alloy and its CPFEM properties evaluation, Journal of Materials Science & Technology 90 (2021) 168-182.
[43] J.-N. Zhou, Y.-F. Guo, J.-Y. Ren, X.-Z. Tang, The mechanism of hcp-bcc phase transformation in Mg single crystal under high pressure, Scripta Materialia 236 (2023) 115670.
[44] S. Wang, T. Wen, J. Han, D.J. Srolovitz, Coherent and semicoherent α/β interfaces in titanium: structure, thermodynamics, migration, npj Computational Materials 9(1) (2023) 216.
[45] Y. Cui, Y. Zhang, L. Sun, M. Feygenson, M. Fan, X.-L. Wang, P.K. Liaw, I. Baker, Z. Zhang, Phase transformation via atomic-scale periodic interfacial energy, Materials Today Physics 24 (2022) 100668.
[46] M. Li, X. Min, K. Yao, F. Ye, Novel insight into the formation of α″-martensite and ω-phase with cluster structure in metastable Ti-Mo alloys, Acta Materialia 164 (2019) 322-333.
[47] D. Sheppard, P. Xiao, W. Chemelewski, D.D. Johnson, G. Henkelman, A generalized solid-state nudged elastic band method, The Journal of Chemical Physics 136(7) (2012) 074103.
[48] L. Gao, X. Ding, T. Lookman, J. Sun, E.K.H. Salje, Metastable phase transformation and hcp-ω transformation pathways in Ti and Zr under high hydrostatic pressures, Applied Physics Letters 109(3) (2016) 031912.
[49] W.G. Burgers, On the process of transition of the cubic-body-centered modification into the hexagonal-close-packed modification of zirconium, Physica 1(7) (1934) 561-586.
[50] G. Kresse, J. Furthmut, Efficient iterative schemes for ab initio total-energy calculations using a plane-wave basis set, PHYSICAL REVIEW B 54 (1996) 1169-1186.
[51] J.P. Perdew, J.A. Chevary, S.H. Vosko, K.A. Jackson, M.R. Pederson, D.J. Singh, C. Fiolhais, Erratum: Atoms, molecules, solids, and surfaces: Applications of the generalized gradient approximation for exchange and correlation, Physical Review B 48(7) (1993) 4978-4978.

[52] J.P. Perdew, K. Burke, M. Ernzerhof, Generalized Gradient Approximation Made Simple, Physical Review Letters 77(18) (1996) 3865-3868.
[53] A. Stukowski, Structure identification methods for atomistic simulations of crystalline materials, Modelling and Simulation in Materials Science and Engineering 20(4) (2012) 045021.